\documentstyle[onecolumn]{mn}
\newcommand{\mincir}{\raise -2.truept\hbox{\rlap{\hbox{$\sim$}}\raise5.truept
\hbox{$<$}\ }}
\newcommand{\magcir}{\raise -2.truept\hbox{\rlap{\hbox{$\sim$}}\raise5.truept
\hbox{$>$}\ }}
\newcommand{\minmag}{\raise-2.truept\hbox{\rlap{\hbox{$<$}}\raise 6.truept\hbox
{$>$}\ }}
\newcommand{\be}{\begin{equation}}
\newcommand{\ee}{\end{equation}}
\newcommand{\ba}{\begin{eqnarray}}
\newcommand{\ea}{\end{eqnarray}}
\newcommand{\brr}{\begin{array}}
\newcommand{\err}{\end{array}}
\newcommand{\bc}{\begin{center}}
\newcommand{\ec}{\end{center}}

\newcommand{\bv}{\mbox{\bf v}}

\newcommand{\bx}{\mbox{\bf x}}

\def\ifm#1{\relax\ifmmode#1\else$\mathsurround=0pt #1$\fi}
\def\kms{\ifmmode\,{\rm km}\,{\rm s}^{-1}\else km$\,$s$^{-1}$\fi}
\def\hmpc{\,h\ifm{^{-1}}{\rm Mpc}}
\def\divv{{\bf \nabla \cdot v}}

\def\fig #1, #2, #3 {
  \smallskip
  \centerline{\psfig{figure=#1,height=#2 in,width=#3 in}} }

\def\\{\hfill\break}

\def\iras{{\it IRAS}}

\def\betac{\beta_c}
\def\delt{\delta_t}
\def\delp{\delta_p}

\def\sigp{\sigma_p}
\def\sigpd{\sigma_{\delta_p}}
\def\sigpv{\sigma_{v_p}}
\def\sigc{\sigma_c}
\def\sigcd{\sigma_{\delta_c}}
\def\sigcv{\sigma_{v_c}}

\def\Ne{N_{eff}}
\def\Nt{N_{tot}}

\def\etal{{\it et al.\ }}
\def\rms{{\it rms\ }}

\def\eg{{\it e.g.}}
\def\ie{{\it i.e.}}

\def\ifm#1{\relax\ifmmode#1\else$\mathsurround=0pt #1$\fi}
\def\kms{\ifmmode\,{\rm km}\,{\rm s}^{-1}\else km$\,$s$^{-1}$\fi} 
\def\hmpc{\,h\ifm{^{-1}}{\rm Mpc}}

\def\ltsima{$\; \buildrel < \over \sim \;$}
\def\lsim{\lower.5ex\hbox{\ltsima}}
\def\gtsima{$\; \buildrel > \over \sim \;$}
\def\gsim{\lower.5ex\hbox{\gtsima}}
 
\def\pmb#1{\setbox0=\hbox{#1}%
 \kern-.025em\copy0\kern-\wd0
 \kern.05em\copy0\kern-\wd0
 \kern-.025em\raise.0433em\box0}
\def\vv{\pmb{$v$}}

\def\v0{\pmb{$0$}}
\def\vnabla{\pmb{$\nabla$}}
\def\div{\vnabla\!\cdot\!}
\def\div{\vnabla\!\cdot\!}

\def\divv{\div\vv}

\title[Cluster versus POTENT fields]
{Cluster versus POTENT Density and Velocity Fields: \\   
  Cluster Biasing and Omega}                             
\author[E. Branchini \etal]
{ E. Branchini$^{1,2}$, I. Zehavi$^{3,4}$, 
M. Plionis$^{5,6}$, \& A. Dekel$^4$ \\ 
$^1$Department of Physics, University of Durham, South Road, Durham DH1
 3LE, UK \\
$^2$Kapteyn Institute, University of Groningen, Landleven 12, 9700 AV,
Groningen, the Netherlands \\
$^3$NASA/Fermilab Astrophysics Group, Fermi National Accelerator
Laboratory, Box 500, Batavia, IL 60510-0500, U.S. \\ 
$^4$Racah Institute of Physics, The Hebrew University,
Jerusalem 91904, Israel  \\
$^5$National Observatory of Athens, Lofos Nimfon, Thesio, 18110 Athens,
Greece \\
$^6$SISSA -- International School for Advanced Studies,
via Beirut 2--4, I--34013 Trieste, Italy }

\begin{document}

\maketitle

\begin{abstract}
The density and velocity fields as extracted from the Abell/ACO
clusters are compared to the corresponding fields recovered by 
the POTENT method from the Mark~III peculiar velocities of galaxies.
In order to minimize non-linear effects and to deal with ill-sampled 
regions we smooth both fields using a Gaussian window with radii 
ranging between $12 - 20\hmpc$. The density and velocity fields within 
$70\hmpc$ exhibit similarities, qualitatively consistent with gravitational 
instability theory and a linear biasing relation between clusters and mass. 
The random and systematic errors are evaluated with the help of mock
catalogs. Quantitative comparisons within a volume containing $\sim\!12$ 
independent samples yield $\betac\equiv\Omega^{0.6}/b_c=0.22\pm0.08$, 
where $b_c$ is the cluster biasing parameter at $15\hmpc$. If $b_c
\sim 4.5$, as indicated by the cluster correlation function, our
result is consistent with $\Omega \sim 1$.
\end{abstract}




\section{Introduction}
\label{sec:intro}

The basic hypothesis underlying the study of large-scale structure is
that it grew out of initial fluctuations via gravitational instability
(GI). In the linear regime, this theory predicts a relation between
the peculiar velocity and density fluctuation fields, $\divv =
-f(\Omega) \delta $, with $f(\Omega) \simeq \Omega^{0.6}$.
From observations we can deduce the density field of galaxies or clusters 
rather than the density field of the underlying matter distribution.
One then needs to assume a relation between the galaxy or cluster fluctuation 
field and that of the mass. A first order approximation is that of a 
linear ``biasing" relation (hereafter LB) in which the two fields, smoothed 
on the same scale, obey 
the relation $\delta_o = b_o \delta$. Thus, GI+LB boil down to a simple
relation between observables,
\begin{equation}
\divv = \beta_o \delta_o \ , \ \
\beta_o \equiv \Omega^{0.6} / b_o \ .
\label{eq:dd}
\end{equation}
The density field of the extragalactic objects can be derived from a whole-sky
redshift survey, while the velocity divergence can be reconstructed from
a sample of redshifts and distances inferred by Tully-Fisher-like
distance indicators.  Therefore, combining these data allow a measure of
$\beta_o$, which, subject to some a priori knowledge of the biasing
parameter $b_o$, provide constraints on the cosmological density
parameter $\Omega$. A related analysis, invoking the integral of equation  
(\ref{eq:dd}), can be performed using velocities rather than densities.

The efforts to measure $\beta$ from various data sets using different
methods are reviewed in \eg, Dekel (1994, 1997), Strauss \& Willick
(1995).  The most reliable density-density analysis, incorporating 
certain mildly-nonlinear corrections, is the
recent comparison of the IRAS 1.2 Jy redshift survey and the Mark~III
catalog of peculiar velocities yielding, at Gaussian smoothing of 
$12\hmpc$, $\beta_{\iras}=0.89 \pm 0.12$ (Sigad \etal 1998, PI98;
replacing an analysis of earlier data by Dekel \etal 1993).
An analysis of optical galaxies has provided a somewhat lower value for 
$\beta_{\rm opt}$ (Hudson \etal 1995), in accordance with the expected 
higher biasing parameter for early-type galaxies as demonstrated by their 
stronger clustering (cf. Lahav, Nemiroff \& Piran 1990). 
Recent velocity-velocity comparisons 
typically yield values of $\beta_{\iras} \simeq 0.5 - 0.6 \pm 0.1$ 
(Willick \etal 1997b and references therein, Willick \& Strauss 1998; 
Davis, Nusser \& Willick 1996, da Costa \etal 1998, Branchini \etal 1999).
The main source of uncertainty in the interpretation of the $\beta$
estimates arises from our ignorance concerning the biasing relations.
Fortunately, we do have a handle on the {\it relative} biasing
parameters, based, for example, on the relative amplitudes of the
correlation functions of the different types of objects, which should
scale like $b^2$.  Since different classes of extragalactic objects are 
assumed to trace the same velocity field, one can hope to tighten the 
constraints on $\Omega$ by deriving $\beta$ for several different types 
of objects.

Clusters of galaxies are promising candidates for this purpose because
they are well-defined objects and are sampled quite uniformly to
large distances, much larger than the available galaxy peculiar velocity
samples. The use of the cluster distribution to probe the large-scale
dynamics has been mainly restricted to dipole analyses, where the
predicted velocity at the Local Group (LG) is compared to its observed
motion relative to the CMB frame (Scaramella, Vettolani \& Zamorani
1991; Plionis \& Valdarnini 1991,  PV91;  Plionis \& Kolokotronis 1998).  
It has been found that the directions
of the two dipoles converge when using a large enough sample of clusters 
($> 150\hmpc$), as expected from the assumed global
homogeneity of the cosmological model (and contrary to the finding of
Lauer and Postman 1994, based on their attempt to directly
measure peculiar velocities for clusters). Once the cluster
distribution is properly corrected from redshift to real space, the
corresponding value of $\beta$ derived from the dipole is $\betac
\simeq 0.21$ (Branchini \& Plionis 1995, 1996, BP96; Scaramella
1995a; Branchini, Plionis \& Sciama 1996). This estimate is higher than
the value derived without this correction (Scaramella \etal 1991; PV91),
and is consistent with $\Omega \sim 1$ for cluster biasing parameter 
of $b_c \sim 4-5$ as indicated by the cluster correlation analyses.
The validity of the LB assumption for clusters might be questioned.
In particular, such a large biasing parameter cannot follow the linear
relation in deep underdensities.
However, because of their low number density, clusters 
trace the underlying mass density field with
a large inherent smoothing scale set by their mean separation.
This has the effect of decreasing the density contrast 
and restoring the plausibility of the LB hypothesis over
a large fraction of the volume sampled.

It turns out that the bulk motion, as predicted from the cluster 
distribution with $\beta_c \sim 0.2$ inside a sphere of radius $\sim
50\hmpc$ about the LG, is consistent with that derived directly from
galaxy peculiar velocities (see Dekel 1997, Dekel \etal 1998; Giovanelli 
\etal 1996, 1998a, 1998b). 
However, the estimate of $\betac$ from the dipole at one point,
or from the bulk flow, naturally suffers from severe cosmic scatter
(\eg, Juskiewicz, Vittorio \& Wyse 1990). The cosmic scatter can be
reduced if the comparison is made at several independent points.
Branchini (1995) and Plionis (1995) have attempted to compare
predicted velocities from the cluster distribution to observed
peculiar velocities of groups and clusters from Tormen \etal (1993),
Hudson (1994), and Giovanelli \etal (1997), obtaining again $\betac
\sim 0.2$. These analyses, however, are of limited validity since they 
compare smoothed and unsmoothed velocities.

The purpose of this work is to measure $\betac$ by comparing the
Abell/ACO cluster distribution and the galaxy peculiar velocities of
the 
comprehensive
Mark~III catalog as analyzed by POTENT.  The comparison is done 
alternatively at the density-density level and at the velocity-velocity 
level, and involves a careful error analysis. 
In \S~\ref{sec:potent} we summarize the Mark~III data, the POTENT method 
and the associated errors.  In \S~\ref{sec:clus} we describe the 
reconstruction of the cluster density and velocity fields and the various 
sources of error. In \S~\ref{sec:beta} we perform a quantitative comparisons 
of the cluster and POTENT fields, in order to determine $\betac$.  We 
conclude our results in \S~\ref{sec:conc}.


\section{POTENT Reconstruction from Peculiar Velocities}
\label{sec:potent}

The POTENT procedure recovers the underlying mass-density fluctuation
field from a whole-sky sample of observed radial peculiar velocities.
The steps involved are:
\begin{itemize}
\item[]{(a)} preparing the data for POTENT analysis, including grouping and
    correcting for Malmquist bias,
\item[]{(b)} smoothing the peculiar velocities into a uniformly-smoothed radial
    velocity field with minimum bias,
\item[]{(c)} applying the ansatz of gravitating potential flow to recover
    the potential and three-dimensional velocity field, and
\item[]{(d)} deriving the underlying density field by an approximation to GI
    in the mildly-nonlinear regime.
\end{itemize}
The POTENT method, which grew out of the original method of Dekel,
Bertschinger \& Faber (1990, DBF), is described in detail in Dekel
\etal (1998, D98) and is reviewed in the context of other methods by Dekel
(1997, 1998). 
Farther improvements since DBF have been introduced which we use in the
present analysis. They are discussed in detail by Sigad \etal (1998)


We use the Mark~III catalog of peculiar velocities  (Willick \etal 1995, 
1996, 1997a), which is a careful compilation of several data sets 
consisting of $\sim 3000$ spiral and elliptical galaxies. 
The non-trivial procedure of merging the data sets accounts for
differences in the selection criteria, the quantities measured, the
method of measurement and the TF calibration techniques. The data per
galaxy consist of a redshift $z$ and a ``forward" TF (or $D_n-\sigma$)
inferred distance, $d$. The radial peculiar velocity is then $u=cz-d$.
This sample enables a reasonable recovery of the smoothed 
dynamical fields 
in a sphere of radius $\sim\!50\hmpc$ about the Local Group, extending
to $\sim 70 \hmpc$ in some well-sampled regions.

The POTENT method is evaluated using mock catalogs.
The mock catalogs and the underlying $N$-body simulation are
described in detail in Kolatt \etal (1996, K96).
Here we only stress that a special effort was made to generate a 
simulation that mimic the actual large-scale structure in the 
real universe, in order to take
into account any possible dependence of the errors on the signal.

\subsection{Errors in the POTENT Reconstruction}
\label{subsec:potent_err}

D98 and PI98 demonstrate how well POTENT can do with ideal data of
dense and uniform sampling and no distance errors. The reconstructed
density field, from input that consisted of the exact, G12-smoothed
radial velocities, is compared with the true G12 density field of the
simulation. The comparison is done at grid points of spacing $5\hmpc$
inside a volume of effective radius $40\hmpc$. No bias is introduced
by the POTENT procedure itself and they find a small scatter of
$2.5\%$ that reflects the accumulating effects of small deviations
from potential flow, scatter in the non-linear approximation
and numerical errors.

Using the mock catalogs described by K96, we
want to check and quantify how well the POTENT reconstruction method
works on our sparse and noisy data. Our goal is to eventually compare
the POTENT fields to the density and velocity fields obtained from the
distribution of clusters. Since the clusters are sparse tracers of the
mass, we need to explore also  smoothing radii larger than the G12
(commonly used in POTENT applications), and we check the G15 and G20
cases as well.  The errors due to sparse sampling and nonlinear
effects are expected to be smaller for the larger smoothing scales,
while the sampling-gradient bias may increase.

For each smoothing radius, we execute the POTENT algorithm on each of
the 20 noisy mock realizations of the Mark~III catalog, recovering 20
corresponding density and velocity fields. We will later consider the
individual fields as well as the mean fields averaged over the mock
catalogs.  The error in the POTENT density field at each point in
space, $\sigpd$, is taken to be the \rms difference over the
realizations between $\delp$ and the true density field of the
simulation smoothed on the same scale.  The errors on the smoothed
velocity fields, $\sigpv$, 
are estimated by a similar procedure
from the Y supergalactic component of the mock velocity field.
We evaluate the density and the velocity fields and their errors at
the points of a Cartesian grid with $5\hmpc$ spacing.  In the
well-sampled regions, with the G15 smoothing,
the errors for the density are typically
$\sigpd\!\approx\!0.1\!-\!0.3$ and $\sigpv\!\approx\! 50 \!-\! 250
\kms$, 
but they are much larger in certain regions at large distances.  

The error estimates $\sigpd$ and $\sigpv$ are 
two of the criteria used to exclude the noisy 
regions from the comparison with the clusters.
The third one is the distance from the 4-th neighbouring object
in the Mark III catalog, $R_4$, which provide us with 
a measure of the poor sampling in the parent velocity catalog.
Two more cuts have been applied on the cluster density and velocity 
fields, using the errors   $\sigma_{v_c}$ and $\sigma_{\delta_c}$
obtained from the mock catalogs analysis in 
\S~\ref{subsubsec:clus_err_mock}. 
Furthermore, we only consider objects within $R=70 \hmpc$ and outside the 
Zone of Avoidance ($|b|>20^{\circ}$).
Our last constraint is on the misalignment angle between the cluster and the
POTENT velocity vectors, $\Delta \theta$, which we impose to be smaller than
45$^{\circ}$. The reason for such additional cut is that we are assuming
all along LB as a working hypothesis. This predicts,
for ideal data, that the velocity vectors reconstructed from the clusters'
distribution should be aligned with the velocity vectors of mass deduced   
by POTENT. However, the various random errors and systematics in both 
types of real data analysed here cause deviations from this simple picture. 
In our ``standard comparison volume'' we restrict the comparison only to
points where the velocity vectors of the POTENT and cluster fields are 
broadly aligned with each other. Note that the misalignment constraint may, 
in principle, affect our $\beta$ estimate and therefore it will be dropped in 
some of the robustness tests performed in \S~\ref{subsec:beta_dd} and
\S~\ref{subsec:beta_vv}.
A ``standard comparison volume'' is defined trough the set of 
cuts reported in Table~\ref{t:cut}.
The two cuts $\sigpv$ and $\sigma_{v_c}$ turned out to be ineffective,
in the sense that they are redundant for reasonable choices of the other 
parameters,  and 
were not implemented. The $\sigma_{\delta_c}$ constraint is obtained by 
scaling  $\sigpd$ by the $\beta_c$ of BP96.
For the sake of consistency, we perform all tests
with the mock catalogs using the same standard cuts, even though the 
$\Delta \theta$ cut does not affect the $\delta$-$\delta$ comparison.
The standard volume, $V_{st}$, depends on the smoothing applied.
For a Gaussian filter of 15 \hmpc has an effective radius 
$R_e \sim 38 \hmpc$ (defined by $(4 \pi/3) R_e^3 = V_{st}$).
The \rms of $\sigpd$ there is $\sim 0.18$ and of $\sigpv$ it
is $\sim 150 \kms$.
Finally, note that the
misalignment criterion depends on the the particular velocity 
field of the generic mock catalog and therefore the same standard cuts 
define slightly different comparison volumes in each of the mock Mark~III
and cluster catalogs tested.


In what
follows we will present the results as the average of the individual
results obtained for each catalog, and for illustrative purposes also
show the results obtained for the mean fields, averaged over the mock
catalogs. (The volumes corresponding to the individual mock catalogs
typically share $80\%$ of their points with the volume defined for the
average fields).
Part of these errors are systematic. The systematic errors can be
evaluated by inspecting  the average of results over the mock catalogs 
or by comparing directly the average POTENT density and velocity fields, 
to the underlying smoothed fields of the simulation. The top panel of
Figure~1 shows this comparison for the G15 smoothed
density fields, at the points of a uniform grid inside the standard
volume.
The residuals in this scatter plot ($\langle \delp \rangle~vs.~\delt$) 
are the local systematic errors.  Their \rms value over the standard volume 
is $0.08$. The corresponding \rms of the random errors ($\delp~ vs.~\langle 
\delp \rangle$) is $0.16$. The systematic and random errors add in 
quadrature to give the total error ($\delp~ vs.~\delt$), whose \rms over the 
realizations at each point, $\sigp$, is used in the analysis below.

To quantify the effect of these errors on the determination of $\beta$
we perform a regression of $\delp$ on $\delt$, for each mock catalog and 
for the average field, by minimizing the following $\chi^2$:
\begin{equation}
\chi^2=\sum_{i=1}^{\Nt}{(\delta_{p,i}-A-B \delta_{t,i})^2\over 
{\sigma_{{\delta_p},i}}^2} \ ,
\label{eq:chi2_p}
\end{equation}
The figure shows no considerable systematic deviations from the $\delp
= \delt$ line (the slope of the regression for the average field comes
out $1.01$). The average of the slopes over the 20 mock realizations
comes out slightly deviant from unity, $1.06$, with a standard deviation
of $0.17$.
For the other smoothings G12 and G20, in the standard comparison
volume, the average of slopes is $1.06 \pm 0.09$ and $0.98 \pm 0.28$,
respectively.  
For other choices of the cuts (see \S~\ref{subsec:beta_dd}
\S~\ref{subsec:beta_vv} for some illustrative examples) 
the slope changes by a couple of percent, \eg, for G15, the typical
deviation from unity is by up to $\sim 5\%$ going either way. 
Finally, only negligible zero point offsets have been detected in all of the
above comparisons. These results indicate that the final systematic
errors are hardly correlated with the signal
or with themselves, 
contributing no significant bias in comparisons 
with density from redshift surveys.

An analogous comparison has been performed between the supergalactic Y
components of the velocity fields at the same points. We limit our 
analysis to this component only as it is the one least affected by the 
uncertainties of the mass distribution near the galactic plane in the 
forthcoming comparison with cluster velocities. Using the two remaining 
Cartesian components would require an estimate of possible systematic 
errors that are uncertain 
for the cluster case (see \S~\ref{subsubsec:clus_err_mock}). 
The bottom panel of Figure~1 shows the corresponding 
scatter diagram of the average field vs. the underlying one, again with 
G15 smoothing in the standard volume.  
The \rms value of the residuals
over the comparison volume which represent the local systematic error, 
($\langle v_p \rangle~ {\rm vs.} ~ v_t$), is $75 \kms$.
The  \rms value of the random errors around the average,
($v_p~{\rm vs.}~\langle~v_p~\rangle$), is $130 \kms$.

Visual inspection of the figure shows clear signs of systematic errors.
The peculiar morphology in the velocity--velocity scatterplot reflects  
correlated velocities within individual cosmic structures.
Indeed, the coherence length of the velocity field is much larger
than for the density field leading to oversampling and correlations  
among the errors. 
The overall effect on the slope is a bias toward smaller values
than unity. The slope of the best-fitting line for the average field 
is in this case $0.81$, and the average of slopes over the mock 
catalogs reflects this as well, giving an average slope of $0.80 \pm 0.18$.
The average slope is $0.93 \pm 0.13$ for the G12 case, and it is $0.76
\pm 0.31$ for G20. When varying the volume, in the G15 case, the bias
is typically $12-22\%$. Thus, the velocity comparisons tend, in
general, to be less robust than the density comparisons and also more
sensitive to the smoothing scale. This is probably partly due to the
larger cosmic scatter in the velocity field, to larger
systematic biases in the POTENT analysis (in particular the window bias 
and the sampling-gradient bias) which become more severe for large 
smoothing scales, 
and to  correlation among the errors.

The POTENT output of the real Mark~III data is similarly provided, for
the three different smoothings, on a Cartesian grid of spacing
$5\hmpc$, within a volume of radius $80\hmpc$. The errors at each grid
point ($\sigpd$ and $\sigpv$) are taken to be the error estimates of
the mock catalogs detailed above, \ie, the \rms difference over the
realizations between the recovered fields and the true underlying one.

\section{Reconstructing the Cluster Density and Velocity Fields}
\label{sec:clus}

The present analysis is based on the real space cluster distribution
and peculiar velocities recovered from the observed distribution of
Abell/ACO clusters in redshift space.  The details of our
reconstruction method are described in BP96.  Here we briefly describe
the data and sketch the main features of the procedure including the
error analysis.  It is worth noticing that in the present comparison
with POTENT we are mainly interested in a local region of radius $\sim
70\hmpc$, where the reconstruction technique is more reliable than at
larger distances.

\subsection{Cluster Data}
\label{subsec:clus_data}

The cluster sample used in BP96 contains all the Abell and ACO
clusters (Abell 1958; Abell, Corwin and Olowin 1989) of richness class
$R \ge 0$ within $250\hmpc$, $|b| \ge 13^{\circ}$ and $m_{10} \le 17$
(where $m_{10}$ is the magnitude of the tenth brightest cluster galaxy
as corrected in PV91). The Abell and ACO catalogs were unified into a 
statistically homogeneous whole-sky sample of clusters using the 
distance-dependent weighting scheme of P91. 
The sample used in this work contains the same $\sim 500$ clusters,
for which $96\%$ now have measured redshifts, most recently from the
ESO Nearby Abell Cluster Survey (Katgert \etal 1996, ENACS). For the
remaining $\sim 20$ clusters the redshifts were estimated from the
$m_{10}$--$z$ relation calibrated as in PV91. 
The results from this improved sample turn out to be fully consistent
with the original ones of BP96.

\subsection{Reconstruction of Uniform Cluster Catalogs in Real Space}
\label{subsec:clus_rec}

Our reconstruction procedure is of two steps. First, Monte Carlo
techniques are used to correct for observational biases and return a
whole-sky distribution of clusters in redshift space. Then, this
distribution is fed into an iterative reconstruction procedure
(similar in spirit to Yahil \etal 1991) which assumes linear GI+LB 
to recover the real-space positions and peculiar velocities of the
clusters.

The main observational biases arise from a systematic mismatch between
the Abell and ACO catalogs and from the latitude-dependent Galactic
obscuration; the radial selection is not an issue because it is quite
uniform in the volume relevant for our analysis. 
To minimize the possible systematic errors in the model cluster velocity 
field we need to  unify the Abell and ACO catalogs into a statistically 
homogeneous whole-sky sample of clusters. We obtain this by using 
the distance-dependent weighting scheme of PV91 which enforces 
the same number density in equal volume shells for the two 
cluster populations. 
The number of radial shells is left as a free parameter.
To correct for Galactic obscuration, we generate a set of
cluster catalogs of uniform sky coverage, by adding a population of
synthetic clusters. The Galactic obscuration at $|b|\geq 20^{\circ}$
is modelled by a cosecant law, ${\cal{P}}(b)$. 
As in BP96 we have chosen two different sets of absorption coefficients
to account for observational uncertainties.
Synthetic clusters are added with a
probability proportional to ${\cal{P}}(b)$ such that they are
spatially correlated with the real clusters according to the observed
cluster-cluster correlation function. Within the ZoA, $|b| <
20^{\circ}$, the volume is filled with synthetic clusters, in bins of
redshift and longitude, by cloning the cluster distribution in the adjacent 
latitude strips outside the ZoA. Both real and synthetic, Monte Carlo 
generated clusters are mass weighted to determine the density field 
to be used in equation (\ref{eq:vv}) below. 
The mass of each real cluster is proportional to the number of galaxies 
per cluster listed in the Abell  catalog. The mass of synthetic clusters 
is set equal to the mass of the real ones.

The redshift space distortions are corrected by an iterative procedure
based on linear theory
and linear biasing. Equation (\ref{eq:dd}) can be inverted to yield 
\begin{equation} 
\bv={\beta_c\over{4\pi}} \int d^3 x'{\delta(\bx')(\bx'-\bx)
\over{|\bx'-\bx|^3}} \,.
\label{eq:vv}
\end{equation}
This is used, in each iteration, to compute the radial peculiar velocities 
of the individual clusters, $u$, 
in the LG frame, 
and correct for their real distances, $r$, via $r=cz-u$.
To avoid strong nonlinear effects, the force field generated by the
point-mass clusters is smoothed by a top-hat window of radius
$15\hmpc$, chosen to be comparable to the cluster-cluster correlation
length (see BP96).  A meaningful comparison with POTENT, in view of
the large mean separation between clusters ($\sim 25\hmpc$), requires
that we smooth further the density and velocity fields.
As input to this procedure one has to assume a value for $\betac$,
which affects the peculiar velocities but has only a weak effect on
the real-space distance as long as $\betac$ is in the right ballpark
(see BP96). We have assumed $\betac=0.21$ based on matching the
dipoles of the CMB and the cluster distribution (\eg, BP96).

\subsection{Errors in the Cluster Positions and Velocities}
\label{subsec:clus_err}

Ideally, we would like to implement the same POTENT error assignment
procedure also for the cluster case. However, intrinsic difficulties
in modelling the Abell/ACO selections criteria hamper the compilation
of mock cluster catalogs from the K96 simulation. A problem which is
made worse by the size of the N-body computational volume which is
smaller than the one spanned by the real clusters' population.  We
choose instead to evaluate cluster errors using a hybrid scheme in
which:
\begin{itemize}
\item The Monte Carlo procedure of adding synthetic clusters and varying
the parameters is also used to estimate the random and systematic
errors that arise both from the uncertainties in modelling the
observational biases and the approximations in the reconstruction.
\item A mock catalog analysis, similar to the one used to assess the POTENT 
errors, is implemented to quantify the additional random errors that
arise from the sparseness of the clusters' sampling.
\end{itemize}

\subsubsection{Monte Carlo Analysis} 
\label{subsubsec:clus_err_mc}

The reconstruction procedure depends on a number of parameters that
are only weakly constrained by observational data or theoretical
arguments, such as the galactic absorption coefficients, the force
smoothing length, and the weighting scheme used to homogenise the
Abell and ACO catalogs.  BP96 evaluated the sensitivity of the derived
density and velocity fields to these parameters by allowing the
parameters to vary about the standard set defined in their Table 2.

The total uncertainties in the cluster positions and 
in the radial velocities, estimated in the CMB frame, 
arise from several different sources:
\begin{itemize}
\item
Intrinsic errors of the reconstruction procedure, which we estimate by
the standard deviation of the cluster distances over 10 Monte-Carlo
realizations of the same choice of parameters.
\item
Observational errors, accounting for the freedom in the values of the
free parameters, which we estimate by the standard deviation of the
distances over reconstructions with different sets of values for the
parameters.
\item
Shot-noise error (eq. [20] of BP96), due to the uncertainty in the mass
per cluster which we assume proportional to the number of galaxies
listed in the Abell catalog (see BP96).  This error is estimated to be
of $\sim 70\kms$ (one dimensional).  The shot-noise due to the
sparseness of the mass tracers will be estimated 
numerically in \S~\ref{subsubsec:clus_err_mock}.  
\item
Weight uncertainty. This error accounts for uncertainties in the
relative weighting of Abell versus ACO clusters to correct for
systematic differences between these catalogs.  For this purpose, we
have performed 10 different reconstructions in which the weights were
randomly scattered about the standard weights of BP96 (their eq. [5]),
following a Gaussian distribution of width that equals the Poisson
error in the relative number densities of Abell and ACO clusters at
every given distance. The estimated typical weight uncertainty turns out 
to be $\sim 85\kms$.
\item
Projection uncertainty. A worry when using the Abell/ACO clusters for
statistical purposes is the contamination of cluster richness due to
projection of foreground and background galaxies (\eg, Dekel \etal
1989 and references therein).
The resulting uncertainty in the galaxy count per cluster has been
recently estimated (Van Haarlem \etal 1997, using N-body simulations;
Mazure \etal 1996 using the ENACS survey) to be $\sim 17\%$.  To
translate this error into a distance error, we have performed 10
reconstructions where the cluster richness were randomly perturbed by
a $17\%$ Gaussian, yielding an error of $\sim 80\kms$.
\end{itemize}

An upper bound to the total error for each Abell/ACO cluster can be
estimated by adding in quadrature all the above errors, as if they are
all independent. This results in an average error of $254\kms$, with a
large spread of $\pm 117\kms$. If add in quadrature only the
observational, intrinsic and shot-noise errors, which are independent
of each other, the average error drops only to $218\kms$, indicating
that our upper bound is not a far over-estimate of the true
error. Figure~2 shows the distribution of the total
reconstruction errors in the line of sight component of the peculiar
velocities for the $\sim 500$ clusters of our subsample.
Unlike the intrinsic ones,
observational, shot-noise, weighting and projection errors are isotropic.
Therefore, they are also representative of 
the uncertainties along the supergalactic Y
component of the velocity fields and they will be used to 
estimate the cluster velocity errors $\sigma_{vc}$ in 
\S~\ref{subsec:clus_fields}).

For the comparison with POTENT we should identify the regions where
the reconstruction from clusters is reliable. The errors are naturally
larger in regions where the fraction of observed clusters is
lower. This effect is clearly seen when we plot the intrinsic error per
cluster as a function of Galactic latitude (Fig. 3).
As expected, no radial dependence has been detected for the errors
within the volume used for the present analysis.

\subsubsection{Mock Catalog Analysis}
\label{subsubsec:clus_err_mock}

Due to their low number density, clusters of galaxies sparsely sample the 
underlying density and velocity fields. This introduces an intrinsic 
scatter, sometimes termed `shot noise' as well, when comparing the 
cluster and the mass fields. 
This is closely related to the expected scatter from the stochasticity   
in the bias relation (\eg Dekel \& Lahav 1999). This sort of random     
errors are not included a-priori in our cluster error estimates, but    
need to be accounted for in the actual comparisons of the clusters       
fields to the POTENT reconstructions.                                      

We further assess the reliability of the cluster fields, and get a
crude estimate of the additional scatter, using the same N-body    
simulation that was the basis for the Mark~III mock catalogs.      
In the present case, however, we obtain just one mock catalog of 
clusters from the 
simulation. We use a friends-of-friends algorithm to identify groups
among the particles in the simulation. The richest groups above some
threshold, fixed so as to have the same number density as the
Abell/ACO clusters, are identified as the mock clusters. 
To mimic the properties of the real cluster distribution we need to extend 
our mock sample out to $250 \hmpc$. Since this exceeds the size of the 
simulation, we obtain the mock cluster distribution by duplicating the 
clusters within the computational box using the periodic boundary conditions. 
The peculiar velocities from clusters are then computed 
from equation (\ref{eq:vv}) and both the cluster density and the velocity
fields are smoothed at the points of a cubic grid with $5 \hmpc$ spacing.
The smoothed cluster fields are then compared with the true underlying
fields of the simulation, smoothed on the same scale. Under the GI+LB
assumptions, the fields are simply related by the biasing factor
between clusters and mass in the simulation, both in the density case
and for the velocities (for our $\Omega=1$ simulation). We use the
same ``standard comparison volume'' considered for the POTENT vs. true
comparisons and used later for the POTENT vs. cluster comparisons. 
Figure~4 shows the results for the G15 case. 

The slopes of the best fitting lines for the $\delta$-$\delta$
comparison ($0.32$, top panel) and for the $v_y$-$v_y$ comparison
($0.30$, bottom panel) 
have been estimated by assigning equal weight to all points
in the plots.
They  are a measure of the relative biasing between
clusters and mass (${b_c}^{-1}$, when regressing clusters on true
fields).  

The average value of the 20 volumes defined by the POTENT
mock catalogs with the same criteria is $0.31 \pm 0.03$ for the
density fields and $0.33 \pm 0.05$ for the velocities.
For general variations in the comparison volume, values of $0.30-0.38$
are typically obtained, with a slight tendency of the values obtained
from velocities to be higher than those obtained from densities,
within this range.
Unlike the POTENT analysis in \S~\ref{subsec:potent_err}, we do not
know the `true' expected slope for the mock clusters and this sort of
comparisons are in practice a way to define it. 
Variations between the values obtained from densities and from
velocities may arise because of the larger cosmic scatter for the
velocities and uncertainties in modelling the cluster distribution
outside the computational volume. 
Another possible cause for the mismatch is the already mentioned 
strong correlation among the errors in the  $v_y$-$v_y$ analysis.
Note that the difference between the slopes obtained from
the $\delta$-$\delta$ and the $v_y$-$v_y$ is smaller than
the outcome of the POTENT analysis in \S~\ref{subsec:potent_err},
meaning the the various error sources affecting the  $v_y$-$v_y$ comparison 
tend to compensate each other.

The distribution of the distant
clusters may significantly affect the cluster velocities while it is
almost irrelevant when computing the smoothed density field within 70
$h^{-1}$ Mpc. We therefore regard ${b_c}^{-1}=0.31$ as the `true'
value for the G15 standard case.  
Similar values are obtained for ${b_c}^{-1}$ with the other smoothing
scales. 

A considerable scatter about the regression lines is found in both the
$\delta$-$\delta$ and the $v_y$-$v_y$ comparisons. Since the clusters
in this case are free from the observational and modelling errors,
this scatter is a manifestation of the additional inherent scatter in 
the cluster fields 
mentioned above.
For G15, the detected scatter is ${\sigma_\delta}^{int}=0.36$ in the
density case and ${\sigma_v}^{int}=300 \kms$ for the velocities. 
The change with smoothing scale is as expected: for G12 the scatter 
is larger (0.46 and 400, for densities and velocities, respectively) 
and for G20 the scatter is smaller (0.24 and 200). These estimates 
are quite robust to changes in the comparison volume.

In what follows, we adopt these dispersions as a measure of the
intrinsic scatter of the cluster fields, for the POTENT-cluster mock
tests in \S~\ref{subsec:beta_test}, and also for the comparisons with
the real data. 
The drawback of the latter assumption is the fact 
that the mock clusters do not accurately match the Abell/ACO cluster 
distribution.  Most of the mock clusters do not correspond on a one-to-one
basis to the Abell/ACO ones, and are less spatially correlated. 
Furthermore, since the mock clusters are identified from a simulation 
based on IRAS galaxies, which
tend to avoid high density regions, they might represent density peaks
of a systematically lower amplitude with respect to those of the
Abell/ACO clusters.  Finally, the a-similarity could be even more
severe for the velocity calculations because of the duplication
procedure adopted outside the N-body computational volume.
Still, not having a better way to accurately determine this scatter
for the real data, and understanding that its magnitude mainly depends
on the sparseness of clusters and smoothing adopted, we believe that
our approach does give a crude estimate of the effect.
It is interesting to check the plausibility our results by comparing them to
the analytic estimates of shot noise computed according to Yahil et al. (1991).
For G15, and within a radius of 70 \hmpc, 
we obtain ${\sigma_\delta}^{an.}=0.42$ for the $\delta$-$\delta$ case
and ${\sigma_v}^{int}=1540 \betac \kms$ for the$v_y$-$v_y$ one.
Scaling to $\betac=0.21$ value of BP96 we obtain that in both 
cases the analytic shot noise is close, although somewhat larger, to 
the scatter in the simulations.
The explicit assumption made here
is that the intrinsic scatter found in the mock simulation is
representative for the real clusters too, and is independent of the
other sources of error and the underlying field. The plausibility of
these hypotheses will be assessed {\it a posteriori} when comparing
the real cluster and POTENT fields.

As for the POTENT analysis (\S~\ref{subsec:potent_err}), the cluster
velocity analysis has been limited to the supergalactic Y component, 
that is less prune to systematics. More extended mock catalog analyses, 
based however on the distribution of IRAS galaxies, have demonstrated 
this point (Branchini \etal 1999). The other two Cartesian components are 
affected by systematic errors, arising from the cloning procedure that 
is used to fill the ZoA. Given the larger extent of the ZoA in the 
cluster case, we expect comparable, if not larger, systematics to affect 
the cluster velocity field. Correcting for this bias would require 
an error analysis based on more realistic Abell/ACO mock catalogs that
are not currently available. Therefore, we restrict our analysis to a 
$v_y$-$v_y$ comparison, under the working hypothesis that the Y component 
of the cluster velocity field is only affected by random errors.

\subsection{Smoothed Density and Velocity Fields}
\label{subsec:clus_fields}

For the purpose of the comparison with POTENT, we compute smoothed
density and velocity fields at the points of a cubic grid with spacing
$5\hmpc$ inside a box of side $320\hmpc$ centered on the Local Group.

We first generate 20 Monte-Carlo realizations of our standard model as
described above. To mimic the effect of observational errors and shot
noise, we perturbed the cluster distances with a Gaussian noise of
$150$ \kms. This value is slightly larger than the sum in quadrature of 
average observational and shot noise errors and 
corrects for the positive tail in the error distribution 
similar to the one observed 
in Figure~2.   
The intrinsic error is not included here because it 
will enter implicitly when we later average over the 20 fields. 
A Cloud-in-Cell (CIC) scheme is used to
translate the discrete cluster distribution into a density field at
the grid points.  The peculiar velocities at the grid points are
recomputed from the reconstructed cluster distribution via linear
GI+LB, with the force field smoothed by a small-scale top-hat window
of radius $5\hmpc$.  We minimize the scale of the top-hat force
smoothing in order to eventually end up as close as possible to
Gaussian smoothing on scales $\geq 12\hmpc$.

The 20 fields are then smoothed further with a Gaussian window of a
larger radius $(R_s^2-R_1^2)^{0.5}$, where $R_1$ is the Gaussian smoothing
radius equivalent in volume to the $5\hmpc$ top-hat force smoothing.
As mentioned already, we try three different smoothing scales, of 
$R_s=12$, $15$ and $20\hmpc$. These 20 fields are averaged to give the 
final smoothed fields used in the next section. 
Each of these 20 fields is affected by observational and shot
noise errors. Moreover, since they represent 20 different 
Monte-Carlo realizations of the cluster distribution for the
same choice of parameters, we can account for the  
intrinsic errors by the very same averaging 
procedure. Indeed, the standard deviation over the 20
realizations represent the cumulative effect of intrinsic, 
observational and shot noise errors discussed in 
\S~\ref{subsubsec:clus_err_mc}. 
The only contribution left to the 
total error budget is the intrinsic scatter estimated in 
\S~\ref{subsubsec:clus_err_mock}. This is modeled as a Gaussian
noise and is added in quadrature at each gridpoint. 
Therefore, the error estimates for the smoothed 
cluster fields, $\sigcd$ and $\sigcv$, are 
obtained by taking the standard deviation over the 20 catalogs
and adding in quadrature a Gaussian noise of amplitude
equal to the intrinsic scatter.

\section{Measuring $\beta_{c}$}
\label{sec:beta}

We determine $\betac$ in two different ways, via $\delta$-$\delta$ and
$v_y$-$v_y$ comparisons. Because the $\delta$-$\delta$ comparison is
local, it avoids the incomplete sky coverage in the ZoA, but it uses
only a small number of clusters (18 within
$70\hmpc$). The $v_y$-$v_y$ comparison, on the other hand, involves an
integral over the cluster distribution in an extended volume, but it
suffers from a large uncertainty due to the unknown cluster
distribution in the ZoA and beyond the sample's edge. Therefore, the
two methods are expected to suffer from different biases
and can provide us with two estimates of $\betac$ that are somewhat
complementary.

We wish to restrict the quantitative comparison to the regions in
space where both the errors in the cluster and POTENT fields are
reasonably small. On the other hand, we wish to maximize the number of
independent volumes compared in order to minimize the cosmic
scatter. We therefore need to optimize our choice of comparison
volume, and test the robustness of the results to changes in this
volume. The comparison volume has already been introduced in 
\S~\ref{subsec:potent_err}.
The natural parameters for defining it the are the
measure of poor sampling of the Mark~III catalog, $R_4$, our estimate
of the random errors in the POTENT and cluster density fields
($\sigpd$ and $\sigcd$) and the corresponding errors in the velocity
fields ($\sigpv$ and $\sigcv$). The latter turned out to be
ineffective cuts, in the sense that they are redundant for reasonable
choices of the other parameters.  Our `standard' cuts according to
these parameters are reported in Table~\ref{t:cut}.
As mentioned already in \S~\ref{subsec:potent_err} we impose as well a
constraint on the misalignment angle between the cluster and POTENT
velocity vectors. This serves as an additional classifier of ``good'' points
for the comparison, and helps avoiding regions where we might have large,
perhaps unaccounted for, errors. 
In our main analysis we restrict the 
comparison to points with a maximal misalignment angle of 
$\Delta \theta < 45^{\circ}$. 
We later relax this constraint and verify 
the robustness of the results. 
We take the G15 smoothing as our standard case, with the above set of 
criteria defining our `standard' comparison volume.

\subsection{The $\betac$ Fitting Method}
\label{subsec:beta_fit}

The assumption underlying the $\betac$ estimations is that the density
and velocity fields recovered above are consistent with the model of
GI+LB. In this framework the POTENT and the cluster fields are
linearly related:
\begin{equation}
p=\betac c+A,
\end{equation}
where $p$ and $c$ stand for the POTENT and cluster and represent
either $\delta$ or the supergalactic Y velocity component.
The cluster errors, $\sigc$, are comparable to the POTENT errors,
$\sigp$. The best-fit parameters are therefore obtained 
by minimising the quantity
\begin{equation}
\chi^2=\sum_{i=1}^{\Nt}{(p_i-A-\betac c_i)^2\over({\sigma_{p,i}}^2+\betac^2
{\sigma_{c,i}}^2)} \ ,
\label{eq:chi2}
\end{equation}
where the subscript $_i$ refers to any of the $\Nt$ gridpoints within
the comparison volume. Since the fields have been smoothed on scales much 
larger then the grid separation, these points are, however, not independent. 
As in Hudson \etal (1995; see also Dekel \etal 1993), we estimate the 
effective number of independent points, $\Ne$, as
\begin{equation}
\Ne^{-1}=\Nt^{-2}\sum_{j=1}^{\Nt}\sum_{i=1}^{\Nt}
\exp(-r_{ij}^2/2R_s^2) \ ,
\label{eq:neff}
\end{equation}
where $r_{ij}$ is the separation between gridpoints $i$ and $j$. This
expression weighs the dependent grid points taking into account
properly the finite comparison volume and its specific shape. This
estimate is thus more accurate than the simplistic ratio of the
comparison volume over the effective volume of the smoothing window,
which assumes an infinite comparison volume. 
We account for the oversampling problem by using an effective $\chi^2$
statistics defined by $\chi^2_{eff} \equiv (\Ne/\Nt) \chi^2$, which is
equivalent to multiplying the individual errors by the square root of
the over-sampling ratio $\Nt/\Ne$. 
The assumption we make is that this new statistics is
approximately distributed like a $\chi^2$ with $\Ne$ degrees of
freedom. In what follows we use it to assess the errors in
$\betac$ and $A$.

\subsection{Testing the Comparison}
\label{subsec:beta_test}

Before performing the comparisons with the real data, we wish to
further quantify the possible systematics that might enter, 
verify the validity of the smoothing scheme adopted and find the
optimal smoothing scale for the comparisons. Also, we would like to
understand whether the intrinsic differences between the density and
velocity field comparisons can affect the results.
We do this by comparing the mock POTENT fields of
\S~\ref{subsec:potent_err} with the mock cluster ones from
\S~\ref{subsubsec:clus_err_mock}. 
The results of the density and velocity comparisons, within the standard 
comparison volume, for each of our three smoothing scales, 
are reported in Table \ref{t:pc}. 
The values quoted in the table are all 
the mean values averaged over the results of the 20 mock
POTENT catalogs.
The G15 case is illustrated in Figure~5, where the average
POTENT fields are compared to the cluster ones. 

The values obtained for $\betac$ in the G15 case are encouragingly
close to the ``true'' value of $0.31$ (\S~\ref{subsubsec:clus_err_mock}). 
The typical range of values obtained when altering the chosen volume
for the comparison, from densities and velocities, lie in the range
$0.26-0.35$.

%
%

There are small differences for the other smoothings, with a tendency
towards smaller $\betac$ for the G20 case.  All variations are,
however, well within the formal $\chi^2_{eff}$ error-bars. In all
cases, the zero-point found is hardly significant and is consistent
with zero, within the error-bars.

The $\chi^2/\Ne$ values (defined as S in the table) are only slightly
smaller than unity for the density comparisons, well within the accepted range 
given the small number of degrees of freedom ($1 \pm 0.4$ for G15),
but they are significantly smaller for the velocities. A similar trend
was also found when comparing the POTENT fields with the true N-body
ones. 
This may indicate that model velocities are more correlated than 
the scaling to  $\Ne$ obtained for the densities. 
If true, then our $\chi^2_{eff}$ distribution 
deviates from that of a $\chi^2$
with $\Ne$ degrees of freedom, resulting in an error overestimate
and in a $\chi^2/\Ne<1$ for velocities.  
Note that  this effect is also associated with the 
application of the alignment
constraint, and without it the value of $\chi^2$ increases somewhat.
So, perhaps for aligned vectors our errors are over estimated.
Another, possibly not exclusive, explanation, which is also suggested 
by the POTENT vs.  N-body fields comparison, is
the existence of systematic errors that do not average to zero. 

The formal error-bars obtained for $\betac$, in all cases, are
significantly larger (by a factor of $\sim 2$) than the spread of
values obtained from the 20 Mark~III mock catalogs. 
Again, this may suggest that our 
effective $\chi^2$ statistic recover the correct slope
but overestimates the errors
on $\betac$ also for the $\delta$-$\delta$ comparison.

In summary, the important conclusion from the comparison of the mock
data is that our method provides a fairly reliable estimate of
$\betac$ with no gross biases. 
The results are fairly robust to changes
in the comparison volume and the smoothing scale.
The reasonable $\chi^2$ values obtained for the densities versus the
too-low values for the velocities, incline us to regard, also here,
the density comparison as the more reliable one. 

Our purpose is to constrain $\betac$ in a meaningful way by comparing
the density and velocity fields extracted from the cluster
distribution with the fields recovered by POTENT from peculiar
velocities. The competing obstacles are the very sparse sampling of
the underlying density field by the clusters, on the one hand, and the
limited volume sampled by peculiar velocities, on the other. The
former dictates the use of a large smoothing scale, because
small-scale structure is not traced properly by the clusters,
while the latter calls for a relatively small smoothing scale, in order to
minimize the cosmic scatter associated with the number of independent
volumes, and the systematic biases in the method.  The rough agreement
between the results of different smoothings is encouraging. Large
smoothing scales, such as the G20 case, are perhaps more pruned to
systematics, and in any case, since they reduce the number of
independent data points, they can only constrain $\betac$ very weakly
with large uncertainties. These considerations, along with the fact that 
it matches the intercluster separation, made us choose the G15 filter as 
our standard for the real data analysis.
As mentioned before, the formal errorbars from the $\chi^2_{eff}$
statistics may overestimate the actual uncertainty in the results, but
we conservatively choose to stick with these.

\subsection{Visual Comparison of Maps}
\label{subsec:beta_qual}

Figures~6 display the G15 density and velocity fields (in
the CMB frame) from the clusters (left) and POTENT (right)
reconstructions of the real data 
in three slices parallel to the Supergalactic plane, within a sphere of 
radius $80\hmpc$ about the Local Group. The clusters' densities
and velocities are scaled by $\betac=0.21$. The heavy line delineates our
standard comparison volume.

The similarity between the two density fields is evident in most
regions.  In both fields, the dominant features are the Great
Attractor (on the left), the Perseus-Pisces supercluster (on the
right), and the great void in between. On the other hand, the Coma
supercluster, seen in the clusters map near $(X,Y) \approx (0,70)$, is
not reproduced at the same position in the POTENT map. Differences are
also seen in the upper-right quadrant of the $Z=-25\hmpc$ plane.

There is also some qualitative agreement between the velocity fields,
but it is less striking. The main features common to the two fields
are the convergences into the Great Attractor and into Perseus
Pisces. The main difference is an additional bulk flow from right to
left for the POTENT field, apparent in the three slices. Another
feature absent in the POTENT field is the infall into Coma seen in the
cluster field. Note that the main discrepancies lie outside the
comparison volume, in regions where the errors are expected to be
large at least in one of the reconstructions. These regions will be
excluded from the quantitative comparison below.

It is also worth noticing that the density--velocity maps for the 
clusters are very similar to those obtained by Scaramella (1995b) from
the same Abell/ACO cluster catalogues but using a somewhat 
different technique.

\subsection{Estimating $\betac$ by a Density Comparison}
\label{subsec:beta_dd}
 
We perform the $\delta-\delta$ comparison within the standard volume.
The errors in the POTENT field have been evaluated in
\S~\ref{subsec:potent_err}, and for clusters we use the error estimates 
of \S~\ref{subsec:clus_err}. The results for the three smoothing radii
are displayed in the first three rows of Table~\ref{t:dv}. For the
preferred G15 case we find $\betac=0.20\pm0.07$, and the best-fit
value stays essentially the same for the other cases (with the
errorbar increasing with the smoothing, due to the smaller number of
effective independent points). No significant zero point offset is
found in any of the cases. The $\delta$-$\delta$ scatterplot is
displayed in Figure~7 for the G15 case. The solid line is
the best-fit from the $\chi^2$ minimization.

We have tried several variants of the comparison volume, in order to
check the sensitivity of our results. Two representative examples are 
reported in the last two rows of Table \ref{t:dv}. 
In the forth case there we have considered the original standard volume 
but with a stricter $R_4$ cut, $R_4< 10 \hmpc$. 
The last column shows the results for the most interesting experiment,
i.e. the one in which the misalignment constraint has been removed.
The results of these tests
all confirm the robustness of the $\betac\simeq0.2$ value.
For the density comparisons, we generally get $\chi^2_{eff}/\Ne \approx 1$, 
indicating a good fit. 
Note, that relaxing the constraint on the misalignment angle  
more than doubles the number of gridpoints considered.

As outlined in \S~\ref{subsubsec:clus_err_mock}, 
the present results have been
obtained assuming that the scatter found in the mock cluster fields is
representative of the intrinsic scatter in the real case and that it
is independent of the other sources of error which form $\sigcd$. The
resulting $\chi^2_{eff}$ values are an indication that 
these are indeed fair assumptions.

\subsection{Estimating $\betac$ by a Velocity Comparison}
\label{subsec:beta_vv}

As already pointed out, it is important to perform the POTENT-cluster
regression for the velocities on the ground of its complementarity with 
the $\delta$ analysis. Also, as we have already discussed in 
\S~\ref{subsubsec:clus_err_mock}, we limit the comparison to the
supergalactic Y component which is the more robust of the components. 
We use the same minimising procedure adopted in for the $\delta$-$\delta$ 
comparison. The results are displayed in the right half of Table~\ref{t:dv}.  
For our standard G15 case the result is now $\betac = 0.25 \pm 0.05$,
somewhat higher than the density case, but still consistent within
the errorbars. The scatterplot for this case is shown in
Figure~8.

As was the case for the $\delta$-$\delta$ comparisons, no significant
offset is detected, and the $\betac$ value is quite robust for the
different smoothing scales and under variations of the comparison
volume (the changes in the resulting $\beta$ are well below the $1
\sigma$ significance level).
Note again the peculiar morphology in the scatterplot, arising from 
the coherency in the peculiar velocities within independent cosmic structure. 
Although it may seem that the POTENT and cluster velocity fields 
differ in a large scale bulk flow component,
more quantitative, volume limited comparisons performed by 
Branchini Plionis and Sciama (1996) and Branchini \etal (1999) 
have shown that the two bulk flows agree in amplitude and direction
for a value of $\betac \approx 0.21$

It is especially interesting to check the effect of removing the
alignment constraint (the fifth case in the table).  This is a
demanding robustness check since it extends the $v_y$-$v_y$ comparison
volume to points for which the velocity vectors can be severely
misaligned.  It is encouraging that, even in this case, the slope of
the best fitting line changes only by 4 \%.
%
The $\chi^2/\Ne$ values lie somewhat below unity for all the cases
explored but for the one in which we have removed 
the alignment constraint. In this las case we obtain 
$\chi^2/\Ne \approx 1$ both for the $\delta$-$\delta$
and $v_y$-$v_y$ comparisons.
A similar behavior 
was also obtained for the mock comparisons (\S~\ref{subsec:beta_test}). 

The errors $\sigma_{\betac}^{v}$ obtained from the real analysis are  
smaller than those obtained from the mock and listed in 
Table~\ref{t:pc}.
Even accounting for the difference in the values of $\beta_c$, the 
two error estimates differ by a factor  of $\approx 2$. 
This mismatch probably arises from the characteristics of the 
mock velocity fields.
Indeed, the small computational box used in the original 
K96 simulation and the constraint of having a vanishing bulk velocity 
on the scale of the box, produce a remarkably quiet velocity field
with a bulk velocity of only 100 \kms already on a scale of 40 \hmpc. 
This velocity field has been used to estimate the POTENT 
and part of the cluster velocity errors. 
Real velocities, however, are larger than the mock ones
and these uncertainties probably underestimates the errors for the real case, 
leading to the smaller $\sigma_{\betac}^{v}$ value listed in Table~\ref{t:dv}.

The $v_y$-$v_y$ comparison described above has been performed in the
CMB reference frame. Predicting velocities from galaxy redshift
surveys is commonly done though in the LG frame, in order to minimize
the influence of mass concentrations from outside the sample
volume. The LG frame might therefore be considered the natural frame
in which to perform comparisons with reconstructed velocities. In our
case, the velocities are reconstructed from the far-extending cluster
catalog, which alleviates the above problem and we regard a CMB
comparison as reliable. Furthermore, performing the comparison in the
LG frame would introduce extra complexities, requiring a somewhat
ad-hoc transformation to a common LG frame for both the cluster and
POTENT velocity fields. As a crude test of the sensitivity of our
results to changes in the framework of reference, we shift both velocity 
fields to the cluster LG frame, as defined by the smoothed cluster velocity 
at the origin (with a reasonable choice for $\betac$).  Alternatively we 
consider the peculiar velocities relative to the central observer of each 
reconstructed velocity field independently. The standard G15 comparison of 
the Y components gives $\betac=0.24\pm 0.05$ and $0.25\pm 0.04$ respectively, 
for these two cases, demonstrating once more the robustness of our result.

\section{Conclusions}
\label{sec:conc}

We have used the smooth matter fluctuation field obtained by applying the 
POTENT machinery to the Mark~III dataset and compared it to the density 
field deduced from the Abell/ACO cluster distribution. A similar comparison 
has also been performed between the reconstructed cluster velocities and 
those from the Mark~III catalog, smoothed on the same scale.  
We have performed a careful error analysis using mock galaxy and cluster
catalogs derived from N-body simulations. The mock catalogs used in our
POTENT error analysis were especially designed to reproduce the Mark~III 
characteristics. Uncertainties in the cluster fields, on the other hand,
were evaluated using a hybrid procedure which extends the Monte-Carlo error 
analysis of BP96 and is complemented with a similar to the POTENT mock 
catalog analysis. 

Cluster and POTENT fields show remarkable similarities within $70 \hmpc$, 
while their major discrepancies are usually confined into regions where 
the cluster or the POTENT reconstructions are known to be unreliable.
Quantitative comparisons between cluster and POTENT fields have been
performed in an attempt to estimate the cluster $\beta$ parameter. The
results are quite robust and for the standard G15 case we find
$\beta_c=0.20 \pm 0.07$ from the $\delta$-$\delta$ regression, and a 
somewhat larger value of $\betac=0.25 \pm 0.05$ from the $v_y$-$v_y$
case. This systematic discrepancy is within the $1 \sigma$
significance level, but it is present in all the cases explored.  We
therefore choose to quote a joint estimate for $\beta_c$ of $0.22 \pm 0.08$.

Some differences between the two values are not unexpected given the
different nature of the comparisons.  A similar regression based on
the mock catalogs showed that some discrepancies do
exist. However, in the mock tests the difference between the two
values was of smaller magnitude and in the opposite direction. The
different trends between the real and mock results could arise from the
different modelling of the mass distribution outside the sampled
regions, which can affect the cluster velocity field.
There are other indications for regarding the $\delta-\delta$
results as being more reliable. The $\chi^2/\Ne$ values for the density 
comparison were around unity, while systematically lower values were 
obtained for the velocities. Also the POTENT velocity field was found in 
the mock catalog analysis to suffer of more biases. 

The present analysis suggests a value of $\betac \simeq 0.20-0.25$
for clusters, in accordance with previous estimates.
The distribution of clusters is expected to be biased with respect to
the distribution of galaxies with a biasing factor $b_{cg} \simeq 3-4$
(\eg, from the different correlation lengths obtained for clusters
and for galaxies; Bahcall \& Soneira 1983, Huchra \etal 1990). 
Peacock and Dodds (1994) find such values for the biasing factors, derived 
from the ratios of power spectra calculated for different datasets. Their 
quoted relative biasing factors for Abell clusters, radio galaxies, 
optical galaxies and IRAS galaxies is $4.5:1.9:1.3:1$, respectively. 
Recent results from a comparison of the cluster density
and velocity fields to the fields recovered from the 
PSC$z$ redshift survey constrain this parameter to $b_{cg} =
4.4 \pm 0.6$, with respect to IRAS galaxies (Branchini \etal 1999).
Joined with our constraint on $\betac$ this implies $\beta_I \sim 1$
with, however, a 1-$\sigma$ uncertainty of $\sim 50$ \%.
Although our analysis cannot provide us with a firm $\betac$
determination, due to the large uncertainties associated with the
$\delta$-$\delta$ and $v_y$-$v_y$ comparisons, it leads toward a value
of $\betac$ which is consistent with an Einstein de Sitter universe
for a reasonable cluster linear bias parameter of $b_c \sim 4.5$.
Our value of $b_c$ is a linear fit to the 
$\delta$-$\delta$ and $v_y$-$v_y$ scatterplots.
Under the assumption of Linear Biasing $b_c$ 
represents the relative biasing of Abell/ACO 
clusters with respect to the underlying mass 
density field. Linear Biasing, however, needs not 
to be a good approximation for clusters since 
the large value of $b_c$ causes the LB hypothesis 
to break down in those regions where $\delta<- b_c^{-1}$.
Methods to measure the degree of nonlinearity 
in the biasing relation have been recently developed 
and applied to galaxy distribution
(e.g. Lemson \etal 1999 and Narayanan \etal 1999,
Sigad, Dekel and Branchini 1999) but not yet to 
clusters of galaxies. 
However, there are indirect evidences of the 
small deviation from linear biasing.
The first one comes from the visual inspection
of the $\delta$-$\delta$ scatterplot 
in Figure~7 which does not deviate 
appreciably from LB expectations (the linear fit).
A more convincing evidence of the small deviations from
LB approximation  is obtained 
by performing the $\delta$-$\delta$ comparison for 
various smoothing  filters. 
Increasing the smoothing length decreases the 
amplitude of density fluctuations and reduces the size of those regions
in which the constraint $\delta<- b_c^{-1}$ cause the LB model to fail.
As shown in Table~\ref{t:pc}, increasing the smoothing radius from 12 
to $20 \hmpc$ doesn't change $\betac^{\delta}$ significantly showing that 
regions where LB does not apply play a little role in our analysis.
As a consequence, and for all  practical purposes, Linear Biasing 
is a good approximation on the scales relevant for our analysis and 
and therefore we can regard $b_c$ as the biasing parameter for Abell/ACO clusters.


\section*{Acknowledgments}
We thank the referee Michael Strauss for his helpful comments and
suggestions. We are grateful to Ami Eldar for his help in applying
POTENT to the mock catalogs to Tsafrir Kolatt for providing the
mock simulation and catalogues and to Yahir Sigad for helpful 
discussions.
MP and EB warmly thank Bepi Tormen
for providing his grouped version of Mark~II catalogue which was used
in some of the preliminary work. EB has been supported by an EEC {\em
Human Capital and Mobility} fellowship and acknowledges the 
hospitality of the Hebrew University, where this 
work has been completed
IZ acknowledge supports by the DOE and the NASA grant 
NAG 5-7092 at Fermilab.


\newpage

\newpage

\section*{}

\begin{table*}
\centering
\caption[]{The Standard Cuts.
Column 1: $\sigpd$, Max. error in POTENT $\delta$; \quad
Column 2: $\sigma_{\delta_c}$, Max. error in Cluster $\delta$; \quad
Column 3: $R_4$, Max. distance from the 4-th neighbouring object in  \hmpc;
 \quad
Column 4: $R$, Max. radius in \hmpc ; \quad
Column 5: $|b|$, Min. galactic latitude in  deg. ; \quad
Column 6: $\Delta \theta$, Max. misalignment in deg. ; \quad}
\vspace{-0.7cm}
\tabcolsep 3pt
\begin{tabular}{ccccccc} \\ \\ \hline
$\sigpd$&$\sigma_{\delta_c}$&$R_4$ & $|b|$ & $R$ & $\Delta \theta $ \\ \hline
0.3 & 1.43 & 13 & 20$^{\circ}$ &70 & 45 $^{\circ}$  \\ \hline
\end{tabular}
\label{t:cut}

\vskip 1truecm

\centering
\caption[]{$\betac$ from the POTENT vs. cluster real data comparisons. 
The symbols and units are the same as in Table~\ref{t:pc}. The first three rows
refer to the standard comparison volume. The last two rows are for
G15, when varying the comparison volume (see discussion in the text).}
\vspace{-0.7cm}
\tabcolsep 3pt
\begin{tabular}{ccccccccccccccccc} \\ \\ \hline
$R_s$ & $\Nt$ & $\Ne$ &$R_e$& $\betac^{\delta}$ & $\sigma_{\betac}^{\delta}$ & 
$A^{\delta}$ & $\sigma_A^{\delta}$ & $S^{\delta}$ && $\betac^{v}$ &
$\sigma_{\betac}^{v}$ & $A^{v}$ & $\sigma_A^{v}$ & $S^{v}$
\\ \hline
12 & 1239 & 11.7 &33& 0.21 & 0.06 & -0.02 & 0.06 & 1.28 && 0.26 & 0.05 & 28 & 59 & 0.75 \\
15 & 1855 & 10.6 &38& 0.20 & 0.07 & -0.04 & 0.07 & 1.04 && 0.25 & 0.05 & 27 & 60 & 0.69 \\
20 & 1537 & 5.0 &36& 0.22 & 0.12 & -0.04 & 0.06 & 1.23 && 0.23 & 0.07 & -4 & 83 & 0.58 \\ 
15 & 1429 & 8.5 &35& 0.19 & 0.07 & -0.04 & 0.07 & 1.00 && 0.26 & 0.06 & 22 & 67 & 0.73 \\ 
15 & 4286 & 19.1 &50& 0.20 & 0.06 & -0.02 & 0.04 & 1.01 && 0.26 & 0.05 & 9 & 48 & 1.06 \\ \hline 
\end{tabular}
\label{t:dv}

\vskip 1truecm

\centering
\caption[]{$\betac$ from the POTENT vs. cluster mocks comparisons.
Column 1: $R_s$, the smoothing radius in $\hmpc$; \quad
Column 2: $\Nt$, the number of gridpoints within the volume; \quad
Column 3: $\Ne$, the effective number of independent volumes; \quad
Column 4: $R_e$, the effective radius in $\hmpc$; \quad
Column 5: $\betac^{\delta}$, $\betac$ from $\delta$-$\delta$; \quad
Column 6: $\sigma_{\betac^{\delta}}$, the $\betac^{\delta}$ dispersion; \quad
Column 7: $A^{\delta}$, zero point offset for $\delta$-$\delta$ ; \quad
Column 8: $\sigma_A^{\delta}$, zero point dispersion; \quad
Column 9: $S^{\delta} \equiv \chi^2_{eff}/\Ne$ from $\delta$-$\delta$; \quad
Column 10: $\betac^{v}$, $\betac$ from $v_y$-$v_y$; \quad
Column 11: $\sigma_{\betac^v}$, the $\betac^v$ dispersion; \quad
Column 12: $A^{v}$,  zero point offset for $v_y$-$v_y$; \quad
Column 13: $\sigma_A^{v}$, zero point dispersion; \quad
Column 14: $S^{v} \equiv \chi^2_{eff}/\Ne$ from $v_y$-$v_y$.}
\vspace{-0.7cm}
\tabcolsep 3pt
\begin{tabular}{cccccccccccccccc} \\ \\ \hline
$R_s$ & $\Nt$ & $\Ne$ & $R_e$ & $\betac^{\delta}$ & 
$\sigma_{\betac}^{\delta}$ & 
$A^{\delta}$ & $\sigma_A^{\delta}$ & $S^{\delta}$ && $\betac^{v}$&
$\sigma_{\betac}^{v}$ & $A^{v}$ & $\sigma_A^{v}$ & $S^{v}$
\\ \hline
12 & 694 & 10.4 &27& 0.31 & 0.10 & 0.00 & 0.07 & 0.93 && 0.27 & 0.10 & -46 & 70 & 0.44 \\
15 & 1644 & 12.4 &37& 0.32 & 0.13 & 0.02 & 0.06 & 0.91 && 0.30 & 0.13 & -51 & 82 & 0.41 \\
20 & 1823 & 7.2 &38& 0.26 & 0.27 & 0.02 & 0.05 & 0.98 && 0.28 & 0.19 & -37 & 111 & 0.40 \\ \hline
\end{tabular}
\label{t:pc}
\end{table*}

\newpage

\section*{Figure Captions}

\noindent
{\bf Figure 1} Systematic errors in the POTENT analysis. The POTENT fields recovered
from the noisy and sparsely sampled mock data are compared with the
``true" G15 fields of the simulation. The comparison is at uniform
grid points within our ``standard comparison volume'' of effective
radius $40\hmpc$. Plotted, in both cases, is the POTENT field averaged
over the 20 realizations.
Top: The POTENT density field vs. the true density
field.  Bottom: The POTENT supergalactic Y-component of the velocity
field vs.  the true velocities in the simulation.

{\bf Figure 2} Histograms of the global random+systematic errors from the Monte Carlo
analysis. The plot shows the frequency of the uncertainties on the
line of sight component of the reconstructed cluster velocities. Units
are $\kms$.

{\bf Figure 3} Environmental dependency of Monte Carlo errors. Dependence of the
intrinsic errors on galactic latitude. Abell/ACO clusters in the
northern galactic hemisphere are shown as filled dots while open dots
represent southern clusters.  

{\bf Figure 4} 
Random errors in the cluster mock analysis. Cluster mock density and
velocity fields are obtained from the clusters identified in the K96
N-body simulation and compared to the underlying fields, all with G15
smoothing.  The comparison is at the same points shown in
Fig. 1 Top: $\delta$-$\delta$ comparison.  Bottom:
$v_y$-$v_y$ comparison. The solid lines corresponds to the average
best-fit lines. The scatter about the fitting lines is an estimate 
of the intrinsic scatter in the cluster fields.  

{\bf Figure 5} 
Comparing the mock POTENT and cluster fields. The averaged mock POTENT
fields of \S~\ref{subsec:potent_err} are compared with mock cluster
ones of \S~\ref{subsubsec:clus_err_mock}, within the standard
comparison volume. 
Top: $\delta$-$\delta$ comparison. Bottom:
$v_y$-$v_y$ comparison.  
The values quoted are the average over the 20 catalogs of $\beta_c$ and of 
its estimated error from the $\chi^2$ fit.
The solid lines represent the best-fits averages.  

{\bf Figure 6}
Density fluctuations and projected velocity field in supergalactic X-Y
planes.  The Mark~III-POTENT case is shown on the right and the
cluster fields on the left, all G15 smoothed. The density contour
spacing is $\Delta \delta = 0.15$, solid contours refer to overdense
regions while dashed contours refer to negative overdensities. The
thick line indicates the $\delta = 0$ contour. The heavy line defines
the standard comparison volume. The length of the velocity vectors
have been drawn on the scale of the plot. The cluster density
fluctuations and velocities are scaled by $\betac=0.21$. Top panel
shows the plane defined by supergalactic $Z=+2500 \kms$, Middle panel
shows {\it the} supergalactic plane of $Z=0 \kms$, and the lower panel
the plane defined by $Z=-2500 \kms$. 

{\bf Figure 7}
POTENT versus cluster G15 density field from the real data, at
gridpoints within the comparison volume. The solid line results from
the linear best fit.

{\bf Figure 8}
POTENT versus cluster G15 velocity field from the real data. Only the
Y supergalactic components at gridpoints within the comparison volume
are considered. The best-fit line is marked as well.


\newpage
\clearpage

\begin{thebibliography}{}
\bibitem[]{A}Abell G.O., 1958, ApJS, 3, 211
\bibitem[]{ACO}Abell G.O., Corwin, H.G., Olowin, R.P., 1989, ApJS, 70, 1
\bibitem[]{BS}Bahcall, N.A., Soneira, R.M., 1983, ApJ, 270, 20
\bibitem[]{B}Branchini E., Ph.D. Thesis, 1995,  International School of 
Advanced Studies, Trieste
\bibitem[]{BP1}Branchini E.,  Plionis M., 1995, in Maurogordato S., 
Balkowski C., Tao C., Tr\^an Thanh V\^an J., eds, Proc. of the Moriond 
Astrophysics Meeting on Clustering in the Universe. Editions Frontieres, 
Gif-sur-Yvette Cedex, p. 277
\bibitem[]{BP2}Branchini E., Plionis M., 1996, ApJ, 460, 569 [BP96]
\bibitem[]{BPS}Branchini E., Plionis M., Sciama D.W., 1996, ApJ, 461, L17
\bibitem[]{pscz}Branchini E., 
Teodoro L., Frenk C., Schmoldt I., Efstathiou G., White S., Saunders S.,
Sutherland W., Rowan-Robinson M., Keeble O., Tadros H., Maddox S., 
Oliver S., 1999, MNRAS, 308, 1
\bibitem[]{N}da Costa L.N., Nusser A., Freudling W., Giovanelli R., 
Haynes M.P., Salzer J.J., Wegner G., 1998, MNRAS, 299, 425
\bibitem[]{DNW}Davis M., Nusser A., Willick J., 1996, ApJ, 473, 22
\bibitem[]{DB}Dekel A., Blumenthal G.R., Primack J.R., Olivier S., 1989, ApJ, 
338, L5
\bibitem[]{DBF}Dekel A., Bertschinger E., Faber S.M., 1990, ApJ, 364, 349 [DBF]
\bibitem[]{PI93}Dekel A., Bertschinger E., Yahil A., Strauss M., Davis M., 
Huchra J., 1993, ApJ, 412, 1 
\bibitem[]{D94}Dekel A., 1994, ARA\&A, 32, 371
\bibitem[]{D97}Dekel, A, 1997, in da Costa L.N., Renzini A., 
eds, Galaxy Scaling Relations. Springer, p. 245
\bibitem[]{D98}Dekel A., 1998, in Dekel A., Ostriker, J.P., eds, Formation of
Structure in the Universe. Cambridge Univ. Press, p. 250
\bibitem[]{bias}Dekel A., Lahav O., 1999, ApJ, 520, 24
\bibitem[]{pot}Dekel A., Eldar, A., Kolatt, T., Yahil, A., Willick, J.A., 
Faber, S.M., Courteau, S., Burstein, D., 1999,  ApJ, 522, 1 [D98]
\bibitem[]{FD}Fisher K., Davis M., Strauss M.A., Yahil A., Huchra J.P., 1993, 
ApJ, 402, 42
\bibitem[]{GD}Ganon, G., Dekel, A., Mancinelli, P., Yahil, A. 1998, {it in 
preparation}
\bibitem[]{G97}Giovanelli R., Haynes M.P., Herter T., Vogt N., da Costa L.N., 
Freudling W., Salzer J.J., Wegner G., 1997, AJ, 113, 22
\bibitem[]{G98a}Giovanelli R., Haynes M.,  Freudling W., da Costa L.,
Salzer J., Wegner G., 1998a, ApJ, 505, L91 
\bibitem[]{G98b}Giovanelli R., Haynes M., Salzer J., Wegner G., da Costa L., 
Freudling W., 1998b, AJ, 116, 2632
\bibitem[]{G96}Giovanelli R., Haynes M.P., Wegner G., da Costa L.N., 
Freudling W., Salzer J.J., 1996, ApJ, 464, L99
\bibitem[]{HH}Huchra, J.P., Henry, J.P., Postman, M., Geller, M.J., 1990, ApJ, 
365, 66
\bibitem[]{H94}Hudson M.J., 1994, MNRAS, 266, 468
\bibitem[]{H95}Hudson M.J., Dekel A., Courteau S., Faber S.M., Willick J.A., 
1995, MNRAS, 274, 305 
\bibitem[]{JVW}Juskiewicz R., Vittorio V., Wyse R.F.G., 1990, ApJ, 349, 408
\bibitem[]{KM}Katgert P., \etal 1996, A\&A, 310, 8 
\bibitem[]{KD}Kolatt T., Dekel A., Ganon G., Willick J.A., 1996, ApJ, 
458, 419 [K96]
\bibitem[]{LNP}Lahav, O., Nemiroff, R.J. \& Piran, T., 1990, ApJ, 350, 119
\bibitem[]{LP}Lauer T.R., Postman M., 1994, ApJ, 425, 418 
\bibitem[]{LS}Lemson G., \etal 1999, {\it in prep.}
\bibitem[]{MK}Mazure A., \etal 1996, A\&A, 310, 31
\bibitem[]{NW}Narayanan V., Weinberg D., Branchini E., Frenk C., 1999, {\it in prep.}
\bibitem[]{PD}Peacock J.A., Dodds S.J., 1994, MNRAS, 267, 1020  
\bibitem[]{PV}Plionis M., Valdarnini R., 1991, MNRAS, 249, 46 [PV91]
\bibitem[]{PK}Plionis M., Kolokotronis, V., 1998, ApJ, 500, 1 
\bibitem[]{P}Plionis M., 1995, in Maurogordato S., Balkowski C., Tao C., 
Tr\^an Thanh V\^an J., eds, Proc. of the Moriond Astrophysics Meeting on 
Clustering in the Universe. Editions Frontieres, Gif-sur-Yvette Cedex, p. 273
\bibitem[]{SVZ}Scaramella R., Vettolani G.,  Zamorani G., 1991, ApJ, 376, L1
\bibitem[]{S}Scaramella R.,  1995a, in Maurogordato S., Balkowski C., Tao C., 
Tr\^an Thanh V\^an J., eds, Proc. of the Moriond Astrophysics Meeting on 
Clustering in the Universe. Editions Frontieres, Gif-sur-Yvette Cedex, p. 257
\bibitem[]{s2}Scaramella R.,  1995b, Aatr. Lett. and Comm., 32, 137
\bibitem[]{PI98}Sigad Y., Dekel A., Eldar, A., Strauss M., Yahil A., 1998, 
ApJ, 495, 516 [PI98]
bibitem[]{}Sigad Y., Dekel A., Branchini, E., 1999, 
ApJ {\it submitted}
\bibitem[]{SW}Strauss M.A., Willick J.A., 1995, Phys Rep., 261, 271
\bibitem[]{TM}Tormen G., Moscardini L., Lucchin F., Matarrese S.,
 1993, ApJ, 411, 16
\bibitem[]{VF}Van Haarlem M., Frenk C. S., White S.D.M., 1997, MNRAS, 287, 817
\bibitem[]{W1}Willick J.A., Courteau S., Faber S.M., Burstein D., Dekel A., 
1995, ApJ, 446, 12 
\bibitem[]{W2}Willick J.A., Courteau S., Faber S.M., Burstein D., Dekel A.,
Kolatt T., 1996, ApJ, 457, 460 
\bibitem[]{W3}Willick J.A., Courteau S., Faber S.M., Burstein D., Dekel A.,
Strauss, M., 1997a, ApJS, 109, 333
\bibitem[]{WSD}Willick J.A., Strauss, M.A., Dekel, A., Kolatt, T., 1997b, ApJ,
486, 629
\bibitem[]{WS}Willick J.A., Strauss, M.A., 1998, ApJ, 507, 64
\bibitem[]{YS}Yahil A., Strauss M.A., Davis M., Huchra J.P., 1991, ApJ, 372, 
380

\end{thebibliography}
\end{document}